\documentclass[aps,prl,twocolumn,floatfix,superscriptaddress,amsmath,amssymb]{revtex4-2}

\usepackage{graphicx}
\usepackage{dcolumn}%
\usepackage{bm}
\usepackage[pdftex]{color}
\usepackage[dvipsnames]{xcolor}
\usepackage[colorlinks,bookmarks=false,citecolor=RoyalBlue,linkcolor=BrickRed,urlcolor=OliveGreen]{hyperref}
\usepackage[mathlines]{lineno}

\newcommand{\parR}[1]{\noindent\textbf{\textit{#1}}\hspace{0.5cm}}
\newcommand{\Z}{{\ensuremath{\mathbb{Z}}}}

\begin{document}

\title{Emergent conformal boundaries from finite-entanglement scaling\\in matrix product states}

\newcommand{\ugent}[0]{Department of Physics and Astronomy, University of Ghent, Belgium}
\newcommand{\ulb}[0]{Center for Nonlinear Phenomena and Complex Systems, Université Libre de Bruxelles, Belgium}
\newcommand{\cambridge}[0]{Department of Applied Mathematics and Theoretical Physics, University of Cambridge, United Kingdom}

\author{Rui-Zhen Huang}
\email{ruizhen.huang@ugent.be}
\affiliation{\ugent}

\author{Long Zhang}
\affiliation{Kavli Institute for Theoretical Sciences and CAS Center for Excellence in Topological Quantum Computation, University of Chinese Academy of Sciences, Beijing 100190, China}

\author{Andreas M. L\"{a}uchli}
\affiliation{Laboratory for Theoretical and Computational Physics, Paul Scherrer Institute, Villigen, Switzerland}
\affiliation{Institute of Physics, \'{E}cole Polytechnique F\'{e}d\'{e}rale de Lausanne (EPFL), Lausanne, Switzerland}

\author{Jutho Haegeman}
\affiliation{\ugent}

\author{Frank Verstraete}
\affiliation{\cambridge}
\affiliation{\ugent}

\author{Laurens Vanderstraeten}
\affiliation{\ulb}

\date{\today}

\begin{abstract}
The use of finite entanglement scaling with matrix product states (MPS) has become a crucial tool for studying one-dimensional critical lattice theories, especially those with emergent conformal symmetry. We argue that finite entanglement introduces a relevant deformation in the critical theory. As a result, the bipartite entanglement Hamiltonian defined from the MPS can be understood as a boundary conformal field theory with a physical \emph{and} an entanglement boundary. We are able to exploit the symmetry properties of the MPS to engineer the physical conformal boundary condition. The entanglement boundary, on the other hand, is related to the concrete lattice model and remains invariant under this relevant perturbation. Using critical lattice models described by the Ising, Potts, and free compact boson CFTs, we illustrate the influence of the symmetry and the relevant deformation on the conformal boundaries in the entanglement spectrum.
\end{abstract}

\maketitle

\parR{Introduction.}%
The past decades have witnessed the successful application of ideas from quantum information theory in quantum many-body physics, providing new insights beyond conventional many-body techniques \cite{Zeng2019}. The central insight concerns the entanglement structure in the low-energy states of correlated quantum many-body systems, summarized in the entanglement area law \cite{Eisert2010} for gapped states or the logarithmic violations thereof in critical systems \cite{Vidal2003, Wolf2006}. Here the entanglement Hamiltonian $H_E$, also called the modular Hamiltonian, plays a pivotal role. In a given quantum many-body state, it arises when considering the reduced density matrix of a subsystem, and is defined as
\begin{equation}
    \rho = \frac{1}{\mathcal{Z}} \exp\left( - 2\pi H_E \right),
\end{equation}
in which $\mathcal{Z}$ is a normalization factor preserving the unit trace of $\rho$. The low-lying spectrum of $H_E$ or \emph{entanglement spectrum} often contains fingerprints of the exotic nature of a given quantum state, and can be used as a diagnostic tool in numerical simulations. Famous examples are the degeneracies in the entanglement spectrum of a state with symmetry-protected topological (SPT) order in one dimension \cite{Pollmann2010, Chen2011} or the universal form of the entanglement spectrum of a fractional quantum Hall state \cite{Li2008}. The entanglement spectrum also directly determines the entanglement entropy in a given state, which has been identified as one of the key quantities for characterizing topological order \cite{Kitaev2006, Levin2006}.

\par For \emph{critical} phases, the formalism of conformal field theory (CFT) has revealed universal properties of entanglement spectra in one-dimensional systems \cite{Vidal2003, Calabrese2004, Calabrese2008}. It was observed that the structure of the entanglement spectrum in critical spin chains is the one of a boundary conformal field theory (BCFT) \cite{Laeuchli2013}, an observation that was later formalized \cite{Cardy2016}. In a more recent work, the effect of introducing a finite gap by a relevant perturbation was also tackled in general terms \cite{Cho2017}. 

\par On the other hand, the formalism of tensor networks \cite{mps_peps_review2021} has proven very fruitful for capturing the correlations in \emph{gapped} systems, because they naturally model the entanglement structure inherent in the low energy states in these systems. In one dimension, gapped states are described by the class of matrix product states (MPS), to the extent that all gapped phases of one-dimensional matter can be fully classified by MPS using group cohomology~\cite{Pollmann2010, Chen2011}. In two dimensions, non-chiral topological order can be classified through the class of projected entangled-pair states \cite{Schuch2010} and tensor network techniques provide a direct way for calculating the entanglement spectra of chiral topological states \cite{Cirac2011, Poilblanc2015}.

\par An injective MPS with finite bond dimension always has a finite correlation length, and can therefore never capture the critical nature of a quantum ground state directly. Instead we have to develop a scaling theory that describes the effect of this finite bond dimension, similar to the theory of finite-size scaling that is used for extracting critical data from, e.g., Monte Carlo simulations or exact diagonalizations on finite clusters. Since a tensor network effectively truncates the amount of entanglement in a quantum state, this gives rise to the theory of \emph{finite-entanglement scaling}. The fundamental idea has always been that simulating a critical system through MPS induces a finite length scale in the system \cite{Nishino1996, Tagliacozzo2008, Pollmann2009}, and that this length scale can be used to perform a scaling analysis. This approach has led to a number of interesting results \cite{Pirvu2012, Stojevic2015, Tirrito2018, Pillay2019, Rams2018, Rader2018, Corboz2018, Vanhecke2019, Eberhardter2023}, making MPS methods very powerful for simulating critical phenomena. Yet, despite these successes, the effect of finite-entanglement scaling on the entanglement spectrum itself has not been addressed.

\par In this work, we use the aforementioned CFT results to shed a new light on this fundamental question. We will simulate critical spin chain models with MPS in the thermodynamic limit directly, for which the only approximation is due to the finite bond dimension. We will argue that this entanglement cutoff induces a relevant perturbation of the critical system, and show that this perfectly explains the observed structure in MPS entanglement spectra. In fact, we will show that we can engineer the form of the perturbation by imposing symmetries on the MPS representation.

\begin{figure*}[tbp]
\centering
\includegraphics[angle=0,scale=0.29]{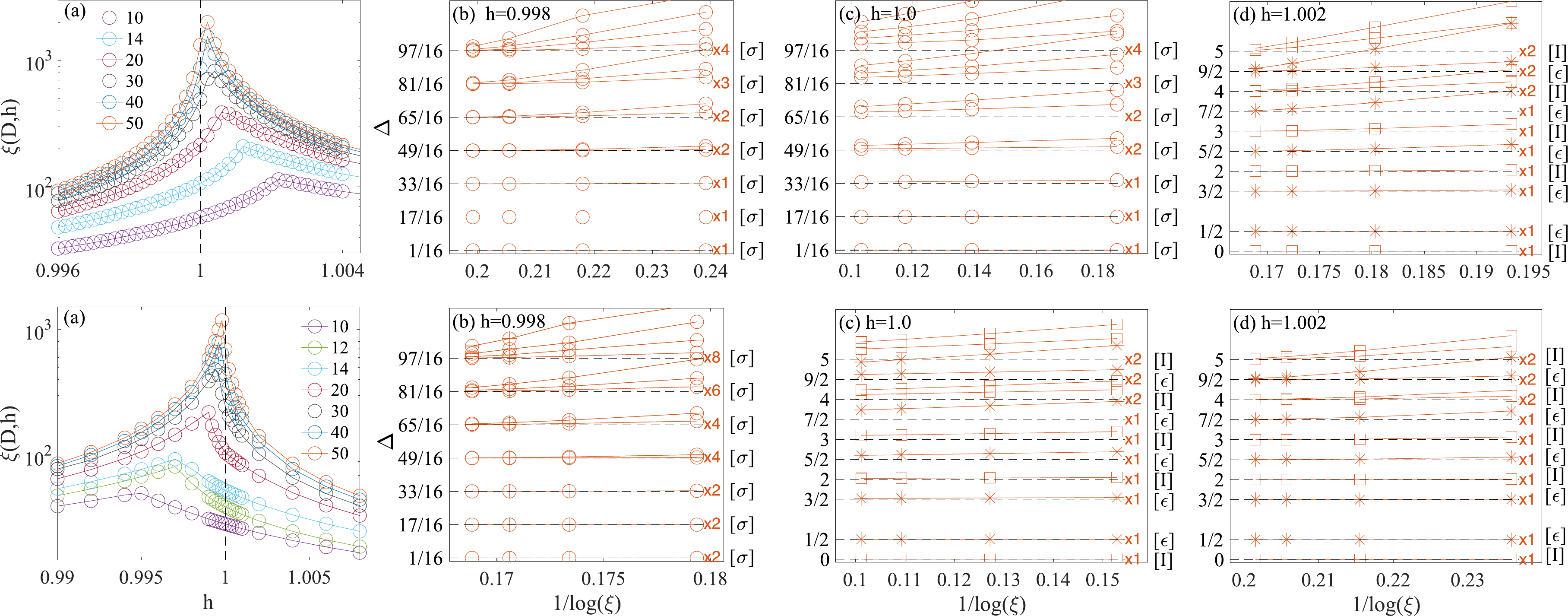}
\caption{Correlation length (a) and entanglement spectra (b-d) for the quantum Ising chain Eq.~\ref{Eq:ising_chain} obtained with non-symmetric MPS (top) and $\Z_2$-symmetric MPS (bottom). In (a) the dashed line denotes the critical point $h=1$ and different curves correspond to different $D$ as shown in the legend. In (b-d) the entanglement spectra have been shifted and rescaled with the first gap. $\sigma$, $I$ and $\epsilon$ represent the spin, identity and energy operators respectively. Dashed lines show the theoretical data. Here and throughout this work, the entanglement spectrum is rescaled according to $\frac{\Delta_i-\Delta_0}{\Delta_1-\Delta_0}h_0$ where $h_0$ is the smallest gap appearing in the conformal family.}
\label{fig_ising}
\end{figure*}

\parR{General framework.}%
Consider a critical spin chain with Hamiltonian $H^*$, for which the low energy properties are described by an effective CFT. We construct a variational MPS ground state approximation for a Hamiltonian in the vicinity of $H^*$ directly in the thermodynamic limit \cite{vmps1, Vanderstraeten2019}, with a given bond dimension $D$. We assume we are working in the regime of finite entanglement scaling where the effective length scale of the MPS state $\xi_D := -1/\log(\vert\lambda\vert) $ ($\lambda$ is the sub-leading eigenvalue of the transfer matrix composed by the MPS local tensor)~\cite{mps_peps_review2021} is much larger than the lattice spacing but smaller than any other length scale (e.g., from not being exactly at criticality). We now propose that this length scale can be modeled as arising from a small relevant perturbation in the CFT. The MPS can thus be viewed as the ground state of a deformed Hamiltonian
\begin{equation} \label{eq_deformed_ham}
    H = H^* + \sum_g \xi_D^{\Delta_g-2}\, O_g + \,\dots \;,
\end{equation}
where the $O_g$ are all relevant operators (operators with a scaling dimension $\Delta_g < 2$) and the dots are additional irrelevant terms. Under a renormalization group (RG) flow, the relevant operators become more and more important. It is the most relevant perturbation $O_g$ governing the scaling properties of MPS. 

\par The effect of this deformation on the entanglement spectrum can be inferred similarly to the discussion in Ref.~\onlinecite{Cho2017}, where the authors considered the imaginary time action of a CFT, perturbed by a primary field. In order to obtain the entanglement Hamiltonian in half-space, one needs to consider a logarithmic conformal mapping, yielding the action 
\begin{equation}
    S = S^* + \int_0^\infty dx \int_0^{2\pi} d\tau \  e^{\left(2-\Delta_g\right)\left(x-\mathrm{log}(\xi_D)\right)} \, O_g + \dots \;,
\end{equation}
where $x$ and $\tau$ denote the Rindler space-time coordinates \cite{Cho2017}, and $O_g$ is the relevant perturbation from Eq.~\ref{eq_deformed_ham}. One can see that the conformal mapping serves the role of an RG process to the relevant deformation. Under the logarithmic mapping, the uniform $O_g$ term becomes exponentially large away from the entanglement boundary and the dynamics is frozen except for the region $ 0 < x < \log\left(\xi_D\right)$. Therefore, it becomes a BCFT with one entanglement boundary $B_e$ and one physical boundary $B_p$, the latter being determined by $O_g$. Significantly, the two boundaries possess distinct physical interpretations and origins. The physical boundary is determined by the relevant deformation in the bulk, while the entanglement one reflects the relation between different degrees of freedom. The entanglement spectrum takes the form of a BCFT spectrum,
\begin{equation}
    \Delta_i \sim \frac{\pi}{\mathrm{log}\left(\xi_D\right)} \left(\Delta_h + n_i \right),
\end{equation}
where $n_i$ is a non-negative integer, $h$ is the primary that marks the conformal tower $V_h$ determined by the fusion rule $V_e \otimes V_g \sim C_{e,g}^h\,V_h$ following the modular invariance of the partition function~\cite{CARDY1984514, CARDY1986200, CARDY1989581,Andrei2020,appendix}. The distribution of $\Delta_i$ provides a clear indicator to identify the deformation $O_g$ to the CFT~\cite{CARDY1984514, CARDY1986200, CARDY1989581,Andrei2020,appendix}. 

Among all the allowed relevant deformations in Eq.~\ref{eq_deformed_ham}, the entanglement truncation in MPS chooses the most relevant perturbation $O_g$ that is allowed by the symmetries of the MPS. In generic MPS (with only translation symmetry), the perturbation imposed by the entanglement truncation should be the primary operator with the lowest conformal weight in the corresponding CFT. In continuous phase transitions with symmetry breaking, the most relevant operator is usually a symmetry-breaking term. This explains why finite entanglement tends to result in ordered MPS at the critical point. However, when we impose symmetry constraints in the MPS representation, such a field term is not allowed in Eq.~\eqref{eq_deformed_ham}, and another deformation will determine the scaling behaviour. So we can use the symmetry constraints in the MPS as a selection rule to impose different physical conformal boundaries in the entanglement spectrum.

\parR{Physical conformal boundaries in MPS.}%
As a first illustration of our approach, we consider the quantum Ising chain with transverse magnetic field
\begin{equation} \label{Eq:ising_chain}
    H = -\sum_i \sigma^z_i \sigma^z_{i+1} - h\sum_i \sigma^x_i .
\end{equation} 
It has a global $\Z_2$ symmetry and undergoes a continuous phase transition when tuning the field across $h=1$. This model realizes an Ising CFT at the critical point. We use variational MPS methods \cite{vmps1, Vanderstraeten2019} to find ground state approximations for a given bond dimension $D$ directly in the thermodynamic limit. We can easily constrain the MPS approximation to be invariant under the $\Z_2$ symmetry by imposing a sparse block structure onto the MPS tensors. We can assume the entanglement boundary to be free~\cite{Laeuchli2013}, but elaborate on this in the next section.
\par In Fig.~\ref{fig_ising}, we show the correlation length and entanglement spectra from non-symmetric (top) and $\Z_2$-symmetric (bottom) MPS simulations. The phase transition is signaled by the singular behaviour of the correlation length $\xi$ \cite{mps_mean_field}, where the singular point approaches the true critical point. An interesting observation is that the direction of this shift of the critical point is different in the $\Z_2$-symmetric and non-symmetric cases. This indeed implies that the symmetry of the MPS representation determines the perturbation $O_g$, and that the scaling behaviour is therefore very different in both cases. 

\par Let us now investigate the entanglement spectra in some detail. First of all, we note that they all show the $1/\mathrm{log}(\xi)$ scaling behavior, in correspondence with the finite-size scaling of Ref.~\onlinecite{Laeuchli2013}. In the symmetry-broken case, the most relevant perturbation is a field term, which induces a fixed up $\vert I\rangle$ or down $\vert \epsilon \rangle$ physical conformal boundary $B_p$.  Together with the free entanglement boundary $B_e$, the entanglement spectrum resembles the operator contents of a BCFT with mixed boundaries, following the fusion rule $\sigma\, \otimes\, I/\varepsilon = \sigma$. Here the fixed or free boundary means a boundary condition without fluctuation in the temporal or spatial direction respectively~\cite{CARDY1984514,CARDY1986200,CARDY1989581,cardy_state1,Blumenhagen:2009zz}. As a result, the spectrum only contains the spin operator family $V_{\sigma}$. In the $\Z_2$-symmetric MPS, the most relevant perturbation is the energy operator, leading to a free physical boundary $\vert \sigma \rangle$. As the entanglement boundary remains free, the entanglement spectrum follows from the fusion rule $\sigma \otimes \sigma = I + \epsilon$. 

\begin{figure}[tbp]
\centering
\includegraphics[width=\columnwidth]{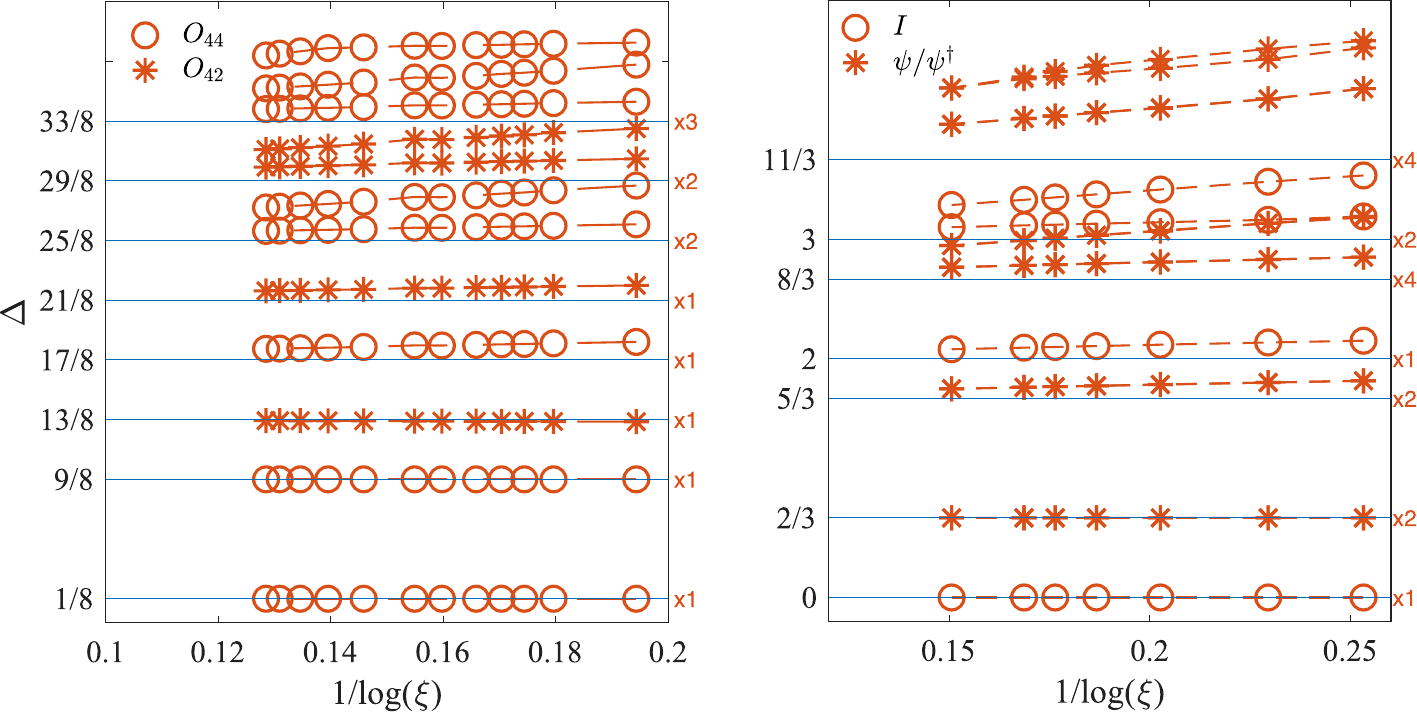}
\caption{Entanglement spectrum obtained from the non-symmetric (left) and $\Z_3$-symmetric (right) MPS for the critical three-state Potts chain. Different markers are used to represent different conformal towers revealed by the spectrum. The spectrum has been normalized with the first energy gap. In the right panel, each $\psi/\psi^\dagger$ spectrum has a two-fold degeneracy (the degeneracy in the two largest ones is slightly lifted at small length scales). $O_{\text{\textit{rs}}}$ represent primaries in the Potts CFT.}
\label{fig:potts}
\end{figure}

A similar analysis also applies to the critical three-state Potts model.~\cite{Ian_Affleck_1998,appendix}, where we can either impose the $\Z_3$ symmetry or not. The result is shown in Fig.~\ref{fig:potts}. One can also consider the charge conjugation symmetry in MPS, where a charge-neutral fixed physical boundary is realized~\cite{appendix}. 

\begin{figure}[tbp]
\centering
\includegraphics[angle=0,scale=0.34]{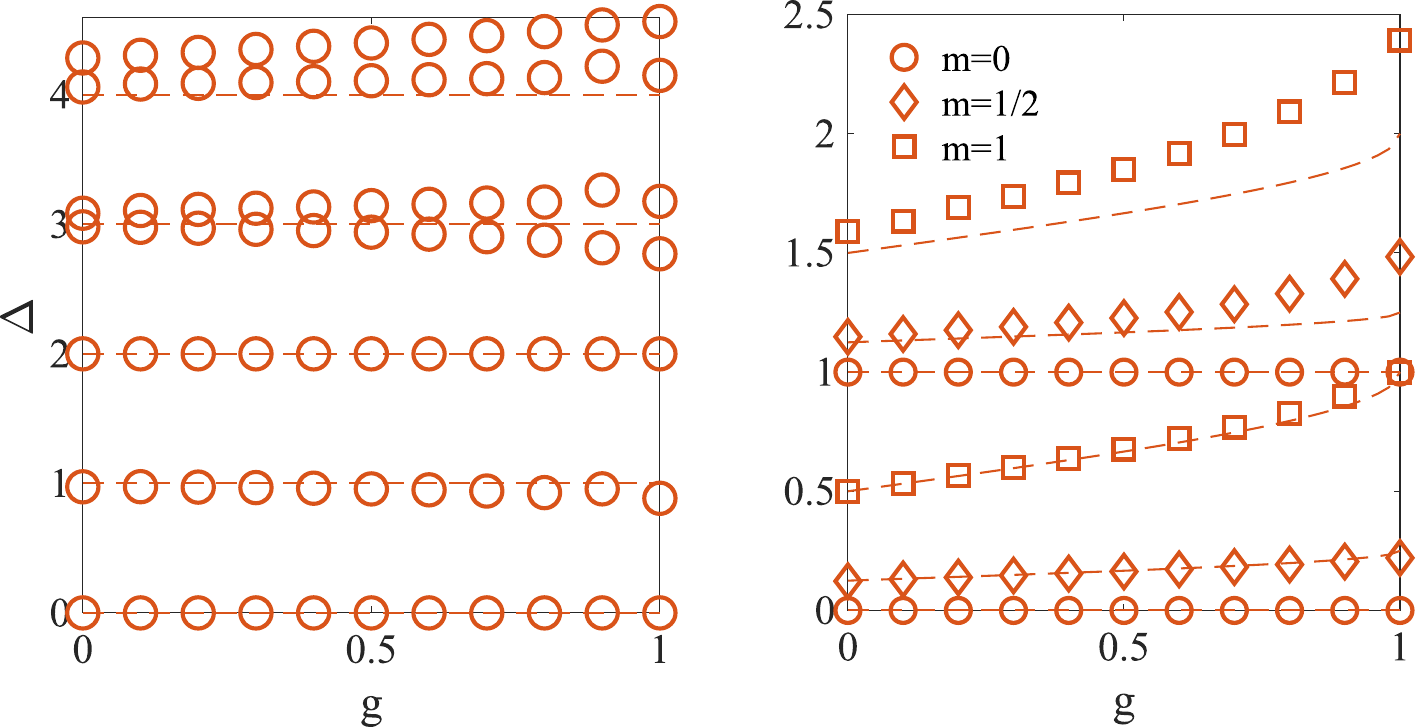}
  \caption{Entanglement spectra for the XXZ spin chain obtained with non-symmetric MPS (left) and $U(1)$-symmetric MPS (right). Dashed lines show the theoretical curve. In the nonsymmetric MPS simulation, we set $D=90$. In the $U(1)$ symmetric MPS simulation, we set the truncation error to be $10^{-5}$. The spectrum on the right is obtained by combining the entanglement spectra from the even and odd cuts in the two-site MPS, without making additional normalizations.}
\label{fig_xxz}
\end{figure}

Besides the minimal models, which describe symmetry breaking transitions, we also study the entanglement spectrum in the XXZ quantum spin chain,
\begin{equation} \label{eq:H_xxz}
    H = \sum_i \sigma_i^x \sigma_{i+1}^x + \sigma_i^y \sigma_{i+1}^y + g\, \sigma_i^z \sigma_{i+1}^z.
\end{equation}
In the region $\vert g \vert <1$, the low energy physics can be described by a $c=1$ free compact boson CFT. When using MPS to approximate its ground state, the finite entanglement cutoff tends to break either the $U(1)$ symmetry resulting in a N\'eel ordered state, or the translation invariance resulting in a dimerized state, as dictated by the Lieb-Schultz-Mattis theorem~\cite{LSM_theorem}. Again, these two cases can be understood as arising from a perturbation of the critical model, in casu, a staggered field in the $xy$ plane or a dimerization term, respectively. The N\'eel ordered case is realized in the non-symmetric MPS simulations, similar to the quantum Ising chain calculation. The dimerized state appears if we use $U(1)$ symmetric MPS with a two-site unit cell, with integer $U(1)$ charges on the odd bonds and half-integer charges on the even bonds.
\par From the abelian bosonization analysis \cite{giamarchi2003quantum,appendix}, the N\'eel and dimerized ordered states correspond to the Dirichlet fixed boundary and the Neumann free boundary conditions for the scalar field $\phi$ respectively. To be consistent with the $U(1)$ symmetry in the state, the entanglement boundary realizes a free boundary condition for the scalar field $\phi$. These boundary conditions can be verified in finite spin chains with a boundary field~\cite{appendix}. As a result, the entanglement spectra are very different in the two cases as shown in Fig.~\ref{fig_xxz}. In the N\'eel ordered case, the entanglement Hamiltonian realizes a mixed boundary BCFT. This case is particularly interesting since the spectrum does not depend on the radius of the compact boson or $g$ in Eq.~\ref{eq:H_xxz} at all. In the dimerized case, the MPS realizes a free boundary BCFT.

\begin{figure}[t]
\centering
\includegraphics[width=\columnwidth]{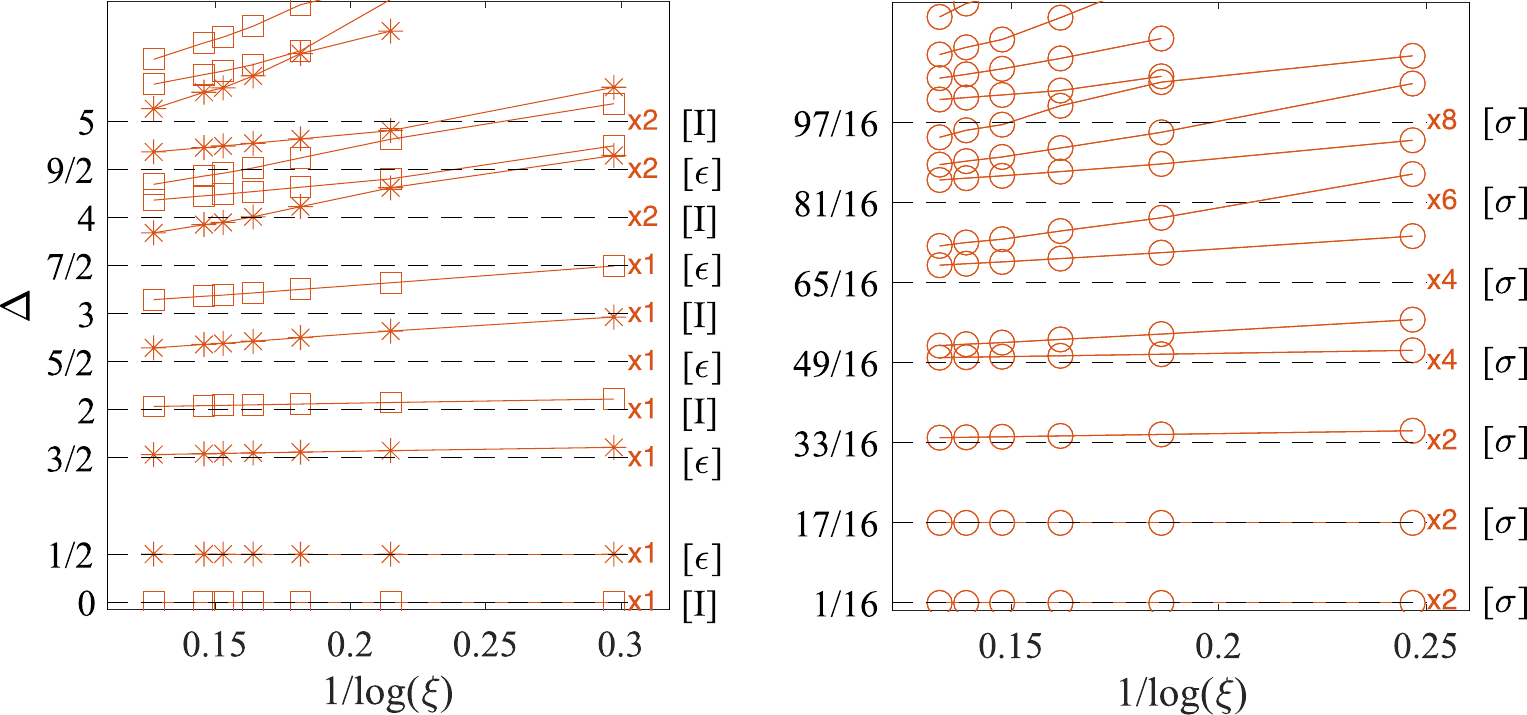}
  \caption{Entanglement spectrum obtained from the non-symmetric (left) and $\mathbb{Z}_2$-symmetric (right) MPS for the critical cluster Ising chain. The entanglement spectrum has been shifted and rescaled with the first energy gap. $\sigma$, $I$ and $\epsilon$ represent the spin, identity and energy operators, respectively. Dashed lines show theoretical values of conformal weights.}
\label{fig_z2k}
\end{figure}

\parR{Entanglement conformal boundary in topological transitions.}%
We have shown how to control the physical boundary $B_p$ in the MPS entanglement spectrum, but the entanglement boundary $B_e$ was fixed by the lattice model. In order to realize non-trivial entanglement boundaries, we can look at transitions between an SPT ordered and a symmetry breaking state \cite{Cho2017b,gapless_spt_2017, symmetry_enrich_ruben_2021, symmetry_enrich_yu_2022}.
\par As an example, we take the cluster Ising spin chain 
\begin{equation} \label{Eq:cluster_ising_chain}
    H = \sum_i - \sigma^z_i \sigma^z_{i+1} - h   \left( \sigma_i^z \tau^x_{i}\sigma^z_{i+1} + \tau_i^z \sigma^x_{i+1}\tau^z_{i+1}\right)\;.
\end{equation}
This model also realizes the Ising CFT at $h=1$ between a symmetry breaking phase and an SPT phase protected by the on-site $\Z_2^\sigma\times\Z_2^\tau$ symmetry. The new term is related to the usual transverse field term in Eq.~\ref{Eq:ising_chain} through a global unitary transformation, which can be viewed as a symmetry twist of the quantum Ising spin chain \cite{gapless_spt_2017, symmetry_enrich_ruben_2021, symmetry_enrich_yu_2022}. This twist results in a different entanglement boundary $B_e$~\cite{symmetry_enrich_yu_2022, appendix}.
\par This is confirmed by the numerical data from the MPS simulation as shown in Fig.~\ref{fig_z2k}, where we show the entanglement spectrum of the nonsymmetric and  $\Z_2$-symmetric MPS at the critical point. Although the MPS share similar bulk properties with the conventional Ising transition in Fig.~\ref{fig_ising}, the operator contents realized in the entanglement spectrum are very different. The physical boundary $B_p$ is the same as in the untwisted Ising model Eq.~\ref{Eq:ising_chain}, but now the entanglement boundary $B_e$ is a superposition of up and down fixed boundary states \cite{gapless_spt_2017,symmetry_enrich_yu_2022}. It appears that the entanglement boundary inherits the nature of the SPT phase transition, even when the bulk states are ordered. As a result, in the symmetric phase the entanglement spectrum has an exact double degeneracy for each level due to the symmetry twisted $B_e$~\cite{appendix}.

\parR{Conclusions and outlook.}%
In this work we have studied the entanglement spectrum in infinite MPS from the BCFT viewpoint. The entanglement Hamiltonian can be described by a BCFT with an entanglement and physical conformal boundary at low energies. We have explicitly related the effect of an entanglement cutoff to a relevant deformation of the CFT describing the critical point. By controlling the symmetry of the MPS, we can alter this deformation and thus the physical boundary. The entanglement boundary, on the other hand, is related to the lattice model and phase transition mechanism. We expect that our work will prove very valuable for extracting universal scaling properties of 1+1 dimensional quantum critical points with tensor networks. 

It would be interesting to extend our insights to more general algebraic symmetries in MPS \cite{Lootens2021, GarreRubio2023} in order to complete all physical boundary conditions. Another interesting question is to identify and classify emergent entanglement boundaries. Since the entanglement boundary remains invariant under bulk deformations, it can be used as an indicator to classify symmetry-enriched quantum critical points. It would also be interesting to study entanglement spectra in critical models without conformal symmetry~\cite{chiral_potts, chiral_clock}: Near a CFT, the conformal symmetry breaking term can be treated as a perturbation and the perturbative picture used here may be applied to study such theories.

Finally, our result has potential applications for studying the entanglement properties of two-dimensional quantum systems. In particular, it is expected that the boundary MPS of critical PEPS \cite{Ran2017, dreyer2020robustness} show a similar scaling of their entanglement spectra \footnote{As an illustrative example, see the Supplemental Material for the case of the nearest-neighbour RVB, which includes Refs.~\cite{ARDONNE2004493,anderson1987resonating,frank_2006}}. Since the boundary MPS is known to encode the entanglement Hamiltonian of the PEPS \cite{Cirac2011}, we can expect that this scaling carries important information on the topological features of the PEPS itself~\cite{appendix}. Also, our results suggest that the finite correlation length scaling in PEPS approximations for two-dimensional quantum critical points \cite{Rader2018, Corboz2018, Vanhecke2019} can also be understood in terms of perturbed 3-D CFT; in particular, we can generalize our framework for engineering the different perturbations arising from imposing symmetry constraints in PEPS.

\parR{Acknowledgements.}%
We would like to thank Shenghan Jiang, Luca Tagliacozzo, Lars Bonnes, Huajia Wang, Laurens Lootens, Maarten Van Damme and Tao Xiang for many helpful discussions. This work was supported by the Research Foundation Flanders, grant no.~G0E1820 and G0E1520N. RZH is supported by a postdoctoral fellowship from the Special Research Fund (BOF) of Ghent University. L.Z. is supported by the National Natural Science Foundation of China (No. 12174387), and the Innovative Program for Quantum Science and Technology (No. 2021ZD0302600). LV is supported by the Research Foundation Flanders (FWO20/PDS/11).

\bibliography{reference}

\newpage\appendix

\section*{Supplemental Material}
In this Supplemental Material, we first give a brief introduction to boundary conformal field theory (BCFT) and its relation to the entanglement spectrum in the ground state of a deformed conformal field theory (CFT). We then present additional results for the entanglement spectrum in the critical Ising, Potts and XXZ spin chains, which correspond to minimal models $\mathcal{M}(4,3)$ ($c=\frac{1}{2}$), $\mathcal{M}(6,5)$ ($c=\frac{4}{5}$) and the free compact boson (c=1) CFTs, respectively. 

\section{A brief introduction to boundary conformal field theory}

A boundary conformal field theory (BCFT)~\cite{CARDY1984514,CARDY1986200,CARDY1989581,Blumenhagen:2009zz} is a CFT defined on a two-dimensional manifold with one-dimensional boundaries. It is an important theoretical tool in the study of surface critical behavior, impurity problems, and open string theory. It is also important in the study of quantum entanglement, since the entanglement Hamiltonian defined with the ground state in a CFT can also be described by a BCFT. Following recent studies of gapped theories in proximity of a CFT~\cite{Cho2017,Cho2017b,Andrei2020}, we have related the entanglement Hamiltonian in MPS to a BCFT with an entanglement and a physical boundary. Here we give a brief introduction to BCFT~\cite{CARDY1984514,CARDY1986200,CARDY1989581,Blumenhagen:2009zz} and its relation to the entanglement spectrum. 

\begin{figure*}[tbp]
\centering
\includegraphics[angle=0,scale=0.42]{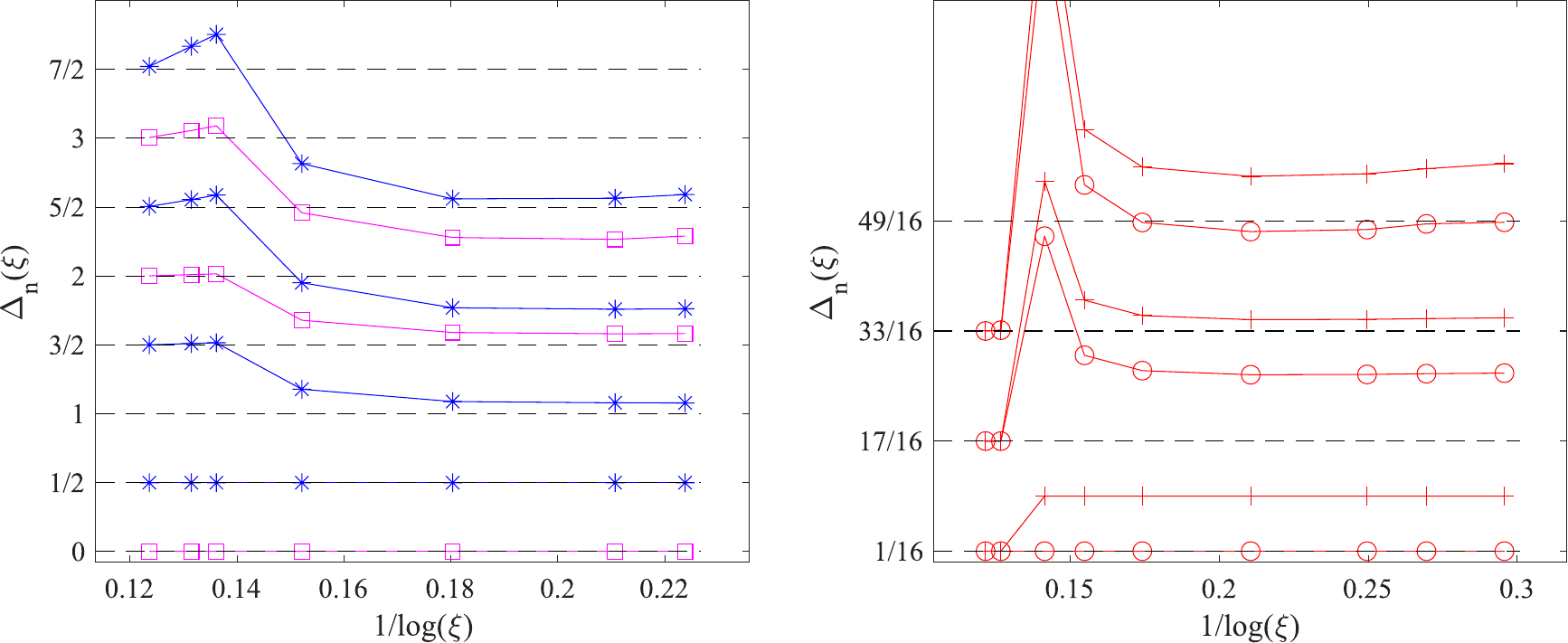}
  \caption{Crossover behavior in the low-lying entanglement spectrum for the critical quantum Ising chain. Left: Entanglement spectrum in the generic MPS at $h=1.0002$. Right: Entanglement spectrum in the $\mathbb{Z}_2$-symmetric MPS at $h=0.9998$. In the $\mathbb{Z}_2$-symmetric MPS, at small length scales the MPS is in the symmetric phase and there is no degeneracy in the entanglement spectrum, so we can use the first gap to normalize the spectrum. While at large length scales, the states are in the symmetry-breaking phase and there is an extra exact two-fold degeneracy, where we use the second gap to normalize the entanglement spectrum.}
\label{fig_cross_over}
\end{figure*}

\subsection{BCFT in the open sector}
A BCFT is defined on a cylinder with two boundaries $B_a$ and $B_b$. The coordinate is denoted by $\omega = \tau + i \sigma$, where we identify the time $\tau \in [0,2\pi \,\beta]$ direction to be periodic and space $\sigma \in [0,\pi]$ to be open. When the time direction range is unbounded $\beta\rightarrow\infty$, the cylinder becomes a strip. Since the Hamiltonian $H_{ab}$ generating time evolution is defined at a fixed time with two boundaries, the whole Hilbert space is constrained by the boundary conditions $B_a$ and $B_b$. One can apply a conformal mapping $z = e^\omega = e^\tau e^{i\sigma}$ to relate the strip geometry to the upper half-plane. The antiholomorphic coordinate $\bar{z}$ in the upper half-plane can be regarded as holomorphic $z^*$ defined in the lower half-plane. As a result, a CFT with boundaries becomes a chiral CFT defined on a plane. It becomes clear now the boundary condition requires the energy-momentum tensor to be real at the boundary, so that
\begin{equation} \label{Eq_T_open}
    T(z) - \bar{T}(\bar{z})\vert_{\sigma=0,\pi} = 0
\end{equation}
or in terms of Laurent series 
\begin{equation}
    \sum_n L_n z^{-n-2} - \sum_n \bar{L}_n \bar{z}^{-n-2}\vert_{\sigma=0,\pi} = 0. 
\end{equation}
One then finds the two Virasoro algebras now reduce to a single algebra under the constraint $L_n = \bar{L}_n$. All states in the quantum theory should satisfy the boundary conditions. Operators are defined within the Hilbert space determined by $B_a$ and $B_b$. The Hamiltonian operator is half of that in the bulk 
\begin{equation}
    H_{ab} = L_0 - \frac{c}{24}. 
\end{equation}
The partition function on the cylinder becomes 
\begin{equation}
    Z_{ab} = \mathrm{Tr}\,\left( e^{-2\pi \beta\, H_{ab}}\right) = \mathrm{Tr}\,\left( q^{L_0-c/24}\right),
\end{equation}
where $q=e^{2\pi i\, \tau_{o}}$ and the modular parameter $\tau_{o} = i\,\beta$. As stated above, the Hilbert space is determined by the boundaries $B_{a/b}$. A boundary operator $\phi_{h}(z)$ appearing in the theory should be consistent with the boundary conditions. The partition function can be written as 
\begin{equation}
    Z_{ab} = \sum_h n_{ab}^h\, \chi_h(q)
\end{equation}
where the non-negative integer $n_{ab}^h$ selects conformal families appearing in the theory and the character $\chi_h(q) = \mathrm{Tr}_h\,\left( q^{L_0-c/24}\right)$.

\subsection{BCFT in the closed sector}
In the open sector the time direction $\tau$ is chosen to be periodic. However, one has the freedom to define a quantum theory along $\tau$ direction and regard $\sigma$ as time. This is the closed sector in BCFT. The Hamiltonian operator now becomes 
\begin{equation}
    H = \frac{1}{\beta} \left( L_0 + \bar{L}_0 -c/12 \right).
\end{equation}
The physical constraint Eq.~\ref{Eq_T_open} now becomes 
\begin{equation}
    z^2\, T(z) - \bar{z}^2\, \bar{T}(\bar{z})\vert_{\tau=0, \pi} = 0
\end{equation}
in the closed sector. Note that due to the choice of time direction, the boundary condition becomes a constraint to the boundary state $\vert a \rangle$ and $\vert b\rangle$. In terms of Laurent modes this is
\begin{equation}
    \begin{split}
        \left(L_n - \bar{L}_{-n}\right) \vert a \rangle = 0.  
    \end{split}
\end{equation}
The basis state of the solution is the so-called Ishibashi state
\begin{equation}
    \vert h \rangle\rangle = \sum_n \vert h,n\rangle \otimes U\vert \bar{h},n\rangle,
\end{equation}
where $U$ is an anti-unitary operator which commutes with the Virasoro generators, $\vert h,n \rangle$ denotes states in a conformal family $\phi_h$. If there are a finite number of representations of the Virasoro algebra in the theory, the physical boundary state (Cardy state) is a superposition of these Ishibashi states
\begin{equation}
    \vert a \rangle = \sum_h \frac{c_a^h}{\sqrt{S_{1h}}}\, \vert h \rangle \rangle,
\end{equation}
where we have imposed the normalization of Ishibashi state according to the modular matrix $S$ defined below.

The same partition function can be rewritten as 
\begin{equation}
    Z_{ab} = \langle a \vert e^{-\pi H} \vert b \rangle = \langle a \vert \,  \tilde{q}^{\frac{1}{2}\left(L_0 + \bar{L}_0 -c/12\right)} \, \vert b \rangle,
\end{equation}
where $\tilde{q}=e^{2\pi i \,\tau_{c}}$ and the modular parameter $\tau_c = i/\beta$. In terms of Ishibashi states, it becomes 
\begin{equation}
    Z_{ab} = \sum_i \frac{\left(c_a^i\right)^* c_b^i} {S_{1i}}\, \chi_i(\tilde{q}).
\end{equation}

\begin{figure*}[tbp]
\centering
\includegraphics[angle=0,scale=0.42]{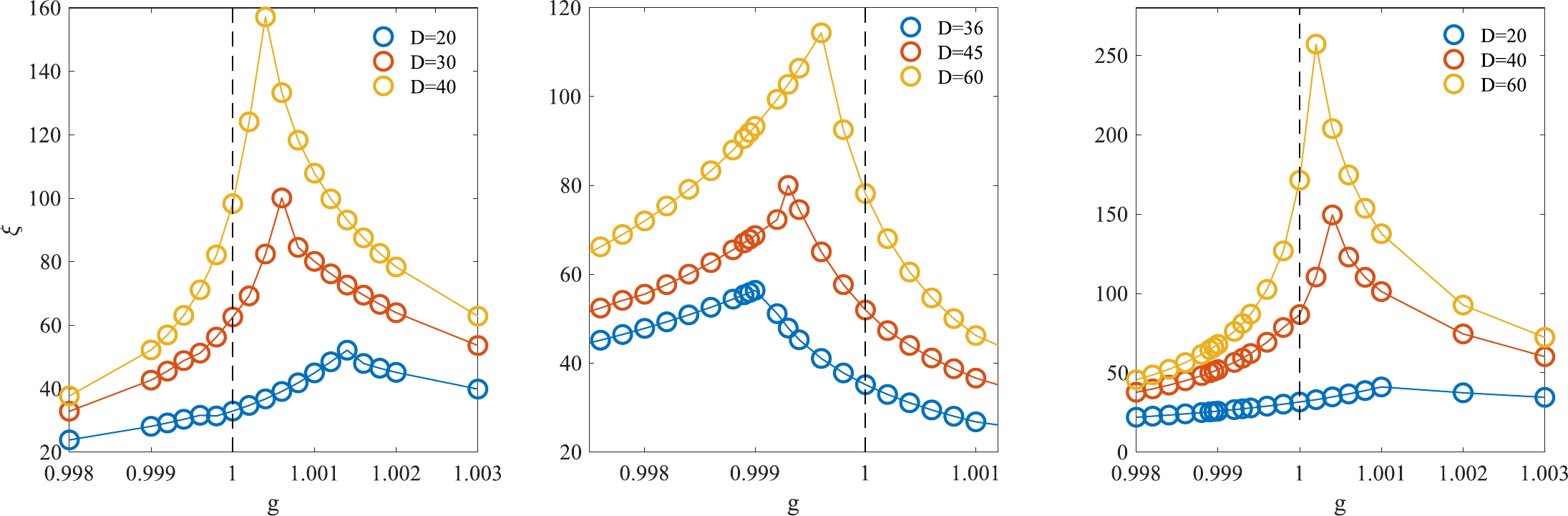}
  \caption{Correlation length in the generic (left), $\mathbb{Z}_3$-symmetric (middle) and charge conjugate $\mathbb{C}$-symmetric MPS for the critical Potts chain Eq.~\ref{potts_ham}. The dashed line shows the critical point $g=1$. In the $\mathbb{Z}_3$ and $\mathbb{C}$-symmetric MPS case, the bond dimension means the total dimension of all charge sectors and each sector has the same dimension.}
\label{fig_potts_xi}
\end{figure*}

\subsection{Open-closed BCFT duality}
The open and closed sector theories are essentially describing the same theory. They are related by a modular transformation $S$ of the modular parameter $\tau \rightarrow -1/\tau$ which exchange the role of space and time. The characters change according to
\begin{equation}
\begin{split}
    & \chi_i(q) = \sum_j S_{ij}\, \chi_j(\tilde{q})\\
    & \chi_j(\tilde{q}) = \sum_i S_{j\bar{i}}\, \chi_i(q)\\
\end{split}
\end{equation}
where $\bar{i}$ denotes the charge-conjugate representation of $i$. This gives the Cardy condition to BCFT
\begin{equation}
\begin{split}
    & n_{ab}^h = \sum_i c_a^i \left(c_b^i\right)^* \, \frac{S_{hi}}{S_{1i}} \\
    & \frac{\left(c_a^i\right)^* c_b^i} {S_{1i}} = \sum_h n_{ab}^h S_{hi}
\end{split}
\end{equation}
where we have used $n_{ab}^h = n_{ba}^{\bar{h}}$.

For self-conjugate minimal models, Cardy states can be constructed by
\begin{equation}
\begin{split}
    c_a^h &= S_{ah}\\
    n_{ab}^h &= \sum_i \frac{S_{ai}\, S_{bi}\, S_{hi}}{S_{1i}}.
\end{split}
\end{equation}
Compared with the Verlinde formula, one can identify the coefficient $n_{ab}^h$ as the fusion multiplicities $N_{ab}^h$. It is now clear that the conformal families appearing in the open sector theory are determined by the fusion of the primaries representing the Cardy states in the closed sector. Note that in this work, we refrain from differentiating between the conformal boundary and conformal boundary state when the context is unambiguous.

\begin{figure}[tbp]
\centering
\includegraphics[angle=0,scale=0.42]{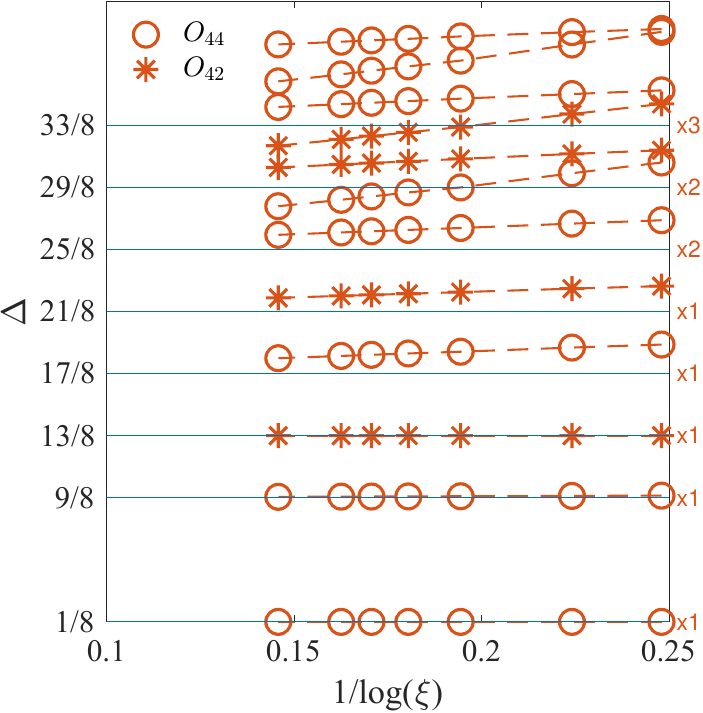}
  \caption{Entanglement spectrum obtained from the charge conjugate symmetric MPS for the critical Potts chain Eq.~\ref{potts_ham}. Different markers are used to represent different conformal towers revealed by the spectrum. The spectrum has been normalized with the first energy gap. Each $\psi/\psi^\dagger$ spectrum has a two-fold degeneracy (the degeneracy in the two largest ones is slightly lifted at small length scales). The first energy gap is used for the normalization.}
\label{fig_potts_dims}
\end{figure}

\section{Entanglement spectrum in uniform infinite MPS}
MPS are essentially gapped theories and they can be fully classified by the symmetry group $G$ and its second cohomology group $H^2(G,U(1))$~\cite{Pollmann2010, Chen2011}. When approximating critical ground states, there is an IR length scale cutoff induced from the entanglement truncation. MPS states can be viewed as ground states of a critical theory deformed by a relevant operator. It is an interesting fact that there is still information of the corresponding critical theory encoded in the entanglement Hamiltonian obtained from the MPS. Close to a CFT, the entanglement Hamiltonian is a BCFT with an entanglement and physical conformal boundaries, in which the physical conformal boundary is determined by the relevant operator that perturbs away from the CFT and opens a gap. The key point here is that we can use symmetry in MPS as a selection rule to determine the most relevant perturbation which results in the symmetric MPS gapped state.

We have assumed that the entanglement boundary is invariant under such a relevant perturbation from the entanglement truncation. In field theories, the entanglement boundary is related to how one regularizes the continuous degrees of freedom. On the lattice it is related to models or interactions between spins. In a topological transition between a SPT and symmetry-breaking state, the entanglement boundary is twisted by symmetries and inherits properties of SPT. These phase transitions are termed as symmetry-enriched quantum critical points. This makes it possible to use different symmetry-enrichment to realize different entanglement boundaries. In these critical points the entanglement boundary usually can be viewed as a superposition of symmetry-breaking boundary states (fixed boundary). The partition function may be written as
\begin{equation}
    Z = \sum_i Z_{E_i,B_g},
\end{equation}
where $i$ labels the degenerate entanglement boundary states.

\subsection{Entanglement spectrum in critical quantum Ising chain}

As discussed above, for CFTs with a finite number of primaries there is a one-to-one correspondence between Cardy states and primaries. In Ising CFT, there are $3$ primaries, the identity operator $I$, spin operator $\sigma$ and the energy operator $\epsilon$. It is a self-conjugate theory and one can use the $S$ matrix to construct the Cardy states~\cite{CARDY1989581}
\begin{equation}
    \begin{aligned}
        \vert I \rangle &= \frac{1}{\sqrt{2}} \vert 0 \rangle \rangle + \frac{1}{\sqrt{2}} \vert \epsilon \rangle \rangle + \frac{1}{2^{1/4}} \vert \sigma \rangle \rangle \\
        \vert \epsilon \rangle &= \frac{1}{\sqrt{2}} \vert 0 \rangle \rangle + \frac{1}{\sqrt{2}} \vert \epsilon \rangle \rangle - \frac{1}{2^{1/4}} \vert \sigma \rangle \rangle \\
        \vert \sigma \rangle &= \vert 0 \rangle \rangle - \vert \epsilon \rangle \rangle.
    \end{aligned}
\end{equation}
They correspond to the up (down), down (up), and free boundaries respectively. Fixed or free physical conformal boundary condition in the entanglement Hamiltonian can be realized by imposing w.r.t a symmetry-breaking or transverse field in the bulk~\cite{cardy_state1}. In the MPS calculations, they correspond the most relevant operator in the generic or the $\mathbb{Z}_2$-MPS respectively. The entanglement spectrum can be predicted according to the fusion of the two primaries labeling the Cardy state.

There also exist interesting cross-over behaviors near the critical point. This stems from the mean-field-like transition in MPS for a given bond dimension $D$. This transition point shifts as we increase $D$ so that there is also a transition at a given parameter near the critical point when enlarging $D$. Figure.~\ref{fig_cross_over} indeed shows there exists crossover behavior from small to large length scales for both the generic and symmetric MPS cases. For the generic case at $h=1.0002$, the entanglement spectrum realizes a mixed boundary BCFT for small $\xi$, while it flows to a free-boundary BCFT for large $\xi$. On the contrary, in the symmetric MPS at $h=0.9998$, the entanglement spectrum flows from a free-boundary BCFT to a mixed-boundary BCFT when enlarging the length scale. These results are consistent with mean-field-like transition behaviors shown in the main text. Note that the cross-over behavior only happens at one side of the critical point since the mean-field-like transition point always shift from one side to the critical point.

In the critical cluster Ising spin chain, the entanglement boundary should be understood as a superposition of the up and down fixed boundaries $\left(\vert I \rangle \oplus \vert \epsilon \rangle\right)$ as analyzed in the main text. The physical boundaries follow a similar analysis for the quantum Ising chain. The MPS state is in the SPT or ordered phases when we impose the $\mathbb{Z}_2$ symmetry or not respectively. As a result, a free or fixed physical boundary is realized accordingly. The entanglement spectrum in the BCFT follow the fusion rule 
\begin{equation}
    \left( I \oplus \epsilon \right) \otimes I/\epsilon = I + \epsilon
\end{equation}
for the non-symmetric and 
\begin{equation}
    \left( I \oplus \epsilon \right) \otimes \sigma = 2\, \sigma
\end{equation}
for the symmetric MPS respectively.

\begin{figure*}[tbp]
\centering
\includegraphics[angle=0,scale=0.42]{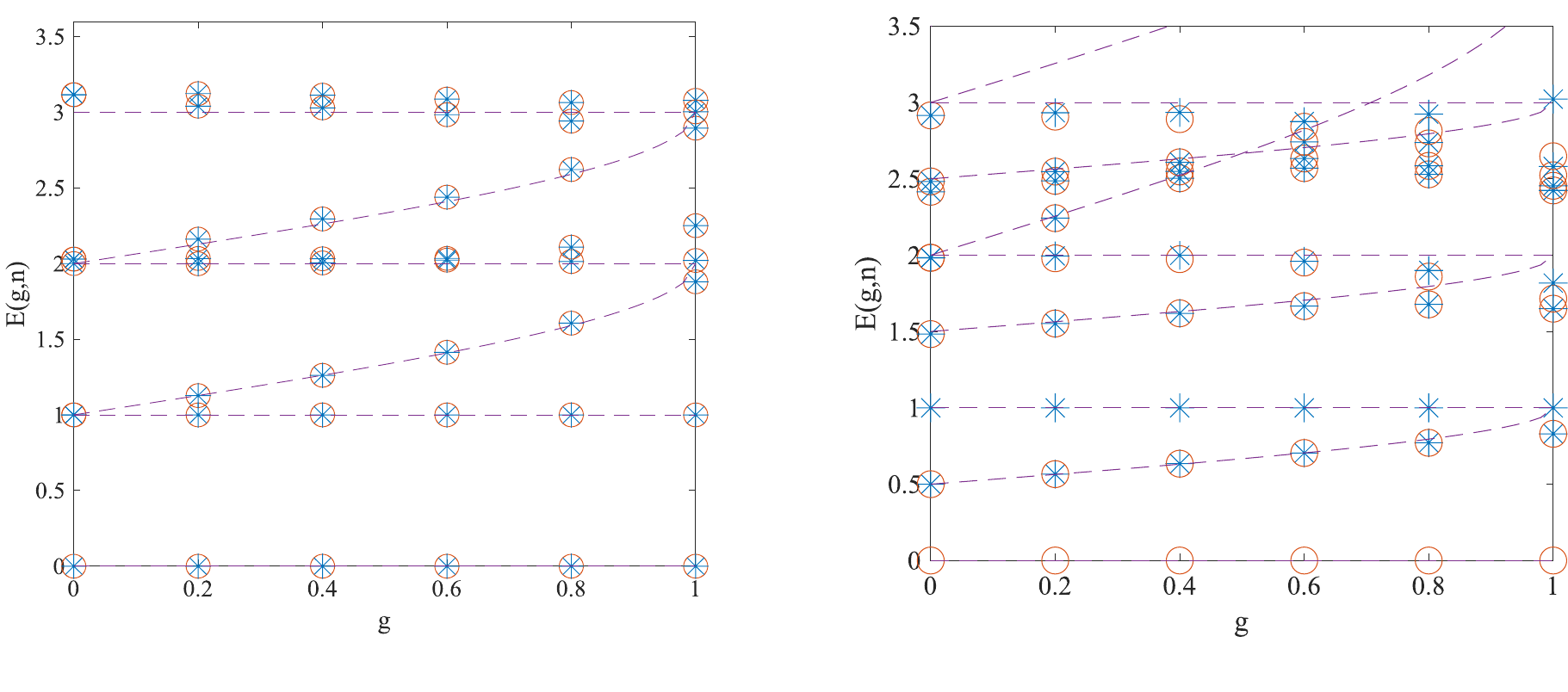}
  \caption{Energy spectra for the XXZ spin chain Eq.~\ref{eq:xxz} with pinned boundaries in the $z$ direction. The low-energy states are obtained from exact diagonalization calculation of a $N=18$ (left) or $N=19$ (right) spin chain. In both cases, the two boundaries are pinned to be up or down at the left or right end respectively in the $z$ direction. Dashed lines show the theoretical curve.}
\label{fig_xxz-e}
\end{figure*}

\begin{figure*}[tbp]
\centering
\includegraphics[angle=0,scale=0.42]{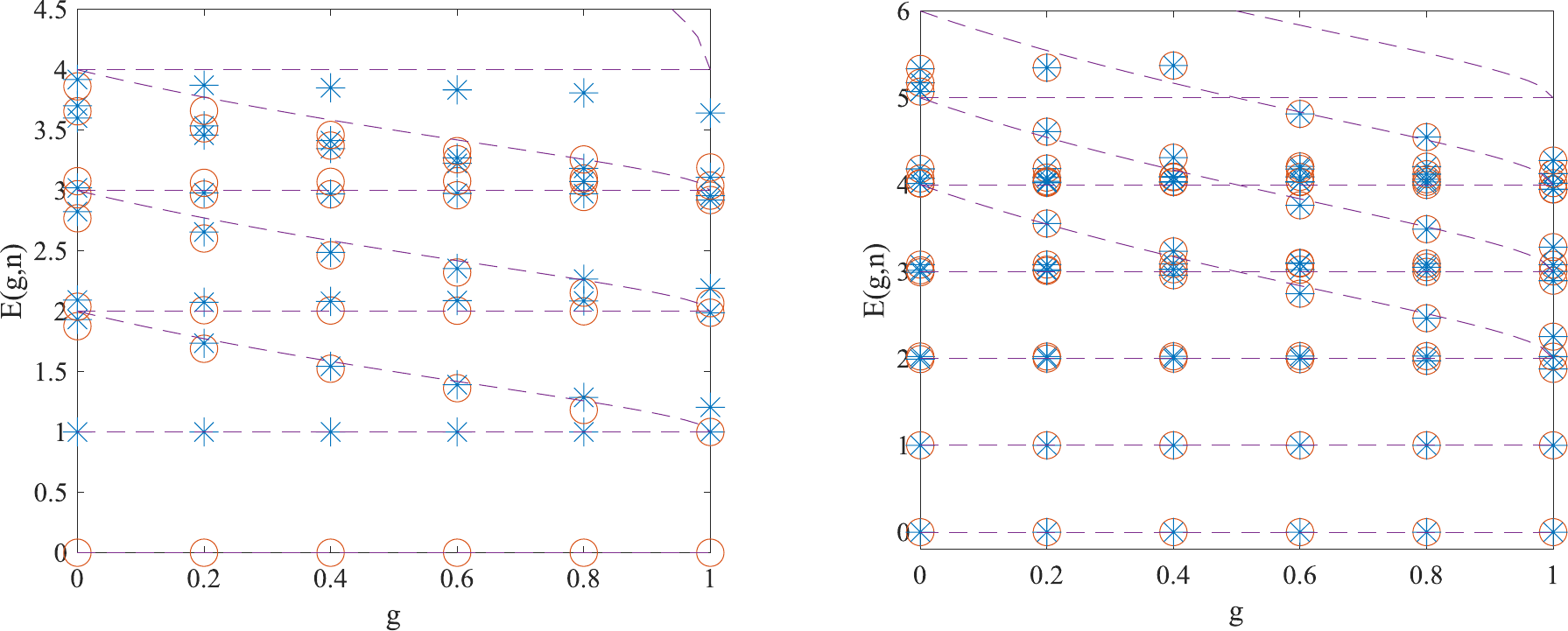}
  \caption{Energy spectra for the XXZ spin chain Eq.~\ref{eq:xxz} with pinned boundaries in the $x$ direction. The low-energy states are obtained from exact diagonalization calculation of a $N=18$ (left) or $N=19$ (right) spin chain. In both cases the two boundaries are pinned to be up or down at the left or right end respectively in the $x$ directions. Dashed lines show the theoretical curve.}
\label{fig_xxz-m}
\end{figure*}

\begin{figure}[tbp]
\centering
\includegraphics[angle=0,scale=0.40]{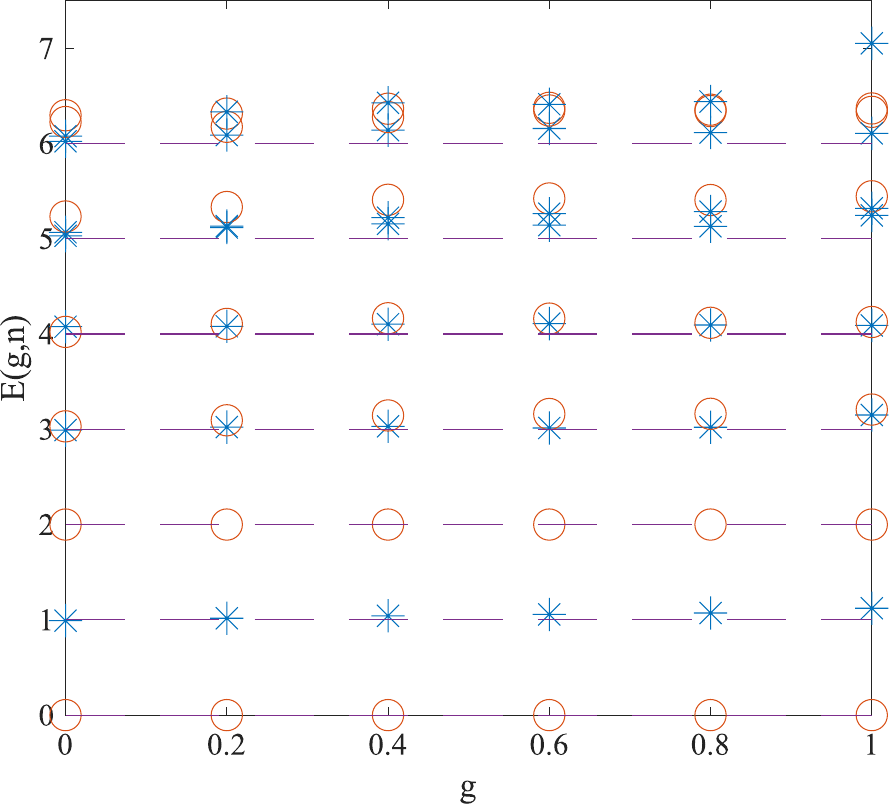}
  \caption{Energy spectra for the XXZ spin chain Eq.~\ref{eq:xxz} with pinned boundaries in the $z$ and $x$ directions at the two ends respectively. The low-energy states are obtained from exact diagonalization calculation of a $N=18$ spin chain.}
\label{fig_xxz-mix}
\end{figure}

\subsection{Entanglement spectrum in the critical quantum three-state Potts model}
A generalization of the quantum Ising spin chain, the three-state Potts spin chain reads
\begin{equation}\label{potts_ham}
    H = \sum_i \left( - Z^\dagger_i Z_{i+1} - g\,X_i  + h.c.\right)
\end{equation}
where $X$ and $Z$ are defined in a $3$ dimensional Hilbert space
\begin{equation}
    X = \begin{pmatrix}
0 & 1 & 0\\
0 & 0 & 1\\
1 & 0 & 0
\end{pmatrix}  
,\, \,
Z = \begin{pmatrix}
1 & 0 & 0\\
0 & \omega & 0\\
0 & 0 & \omega^2
\end{pmatrix}
\end{equation}
with $\omega=e^{i\frac{2\pi}{3}}$. The local Hilbert space can be labeled by an integer $q$ as $\vert q \rangle$ ($q=$0,1,2 mod 3). It has a global $\mathbb{S}_3$ symmetry composed by a $\mathbb{Z}_3$ rotation symmetry which transforms as $q \rightarrow q+1$ and a charge conjugate symmetry $\mathbb{C}$ which transforms as $q \rightarrow -q$.  Its ground state is ordered or symmetric at small or large $g$ respectively. Similar to the quantum Ising spin chain, a duality transformation can map $g$ to $1/g$. At the self-dual point $g=1$, this model realizes the unitary minimal model $\mathcal{M}(6,5)$ with a central charge $c=\frac{4}{5}$~\cite{Ian_Affleck_1998,Blumenhagen:2009zz}. Besides the Virasoro symmetry, the Potts model has an extended $W$-algebra symmetry. The $W$-characters are linear superposition of the Virasoro ones. As a result, there are only four $W$-algebra primaries ($I,\epsilon,\sigma,\sigma^\dagger,\psi,\psi^\dagger$) shown in the modular invariant partition function (Two of them appear twice with opposite $\mathbb{Z}_3$ charges). They form a closed algebra under fusions. 

In the corresponding BCFT, each $W$-primary labels a conformal boundary condition or state. It was known three of them are fixed boundaries ($\vert I \rangle$, $\vert \psi \rangle$, $\vert \psi^\dagger \rangle$) and the other three ($\vert \epsilon \rangle$, $\vert \sigma \rangle$, $\vert \sigma^\dagger \rangle$) are mixed boundary states. It was also found that by fusing these states with other Virasoro-primaries there are two other boundary conditions, the free boundary ($\vert O_{44} \rangle$) and the new boundary ($\vert O_{22}\rangle$). This completes all possible conformal boundary conditions in critical three-state Potts model~\cite{Ian_Affleck_1998}.

The finite entanglement cutoff realizes different scaling functions and relevant perturbations by imposing different symmetries in MPS as discussed in the main text. Without symmetry constraint it is again the symmetry-breaking field (any of the three directions) induced by the entanglement in the bulk. It induces a fixed physical boundary condition. With $\mathbb{Z}_3$ symmetry, the most relevant operator should be the energy operator which results in a $\mathbb{Z}_3$-symmetric state and a free physical boundary. We can also impose the charge conjugate symmetry $\mathbb{C}$ which results in a charge neutral ($q=0$) symmetry-breaking state and a fixed conformal boundary $\vert q=0 \rangle$. Compared with the generic-MPS case, it is the same relevant operator (symmetry-breaking field) induced from the entanglement truncation but resulting in a special fixed boundary ($q=0$).

This analysis is supported by numerical results using MPS. Figure.~\ref{fig_potts_xi} shows the dependence of the correlation length $\xi$ to $g$. In the generic-MPS calculation, the MPS states are ordered at the critical point as a result of the symmetry-breaking field induced by the entanglement cutoff. Similarly in the $\mathbb{Z}_3$ or charge conjugate $C$-symmetric MPS case, the symmetry constraint enforces a symmetric or charge neutral symmetry-breaking state. In the main text we have shown results for the generic and $\mathbb{Z}_3$-symmetric MPS. In Fig.~\ref{fig_potts_dims} we also present the entanglement spectrum from the charge conjugate symmetric MPS. The operator contents appearing in the entanglement spectrum follows from the fusion between primaries labeling the entanglement and physical boundaries. Specifically, in the generic or charge conjugate symmetric MPS case, it is the fusion between free and fixed boundary states
\begin{equation}
    O_{44} \times I = O_{44} + O_{42}.
\end{equation}
and with $Z_3$ symmetry, it is the result of two free boundaries
\begin{equation}
    O_{44} \times O_{44} = I + \psi + \psi^\dagger.
\end{equation}
These conformal families are correctly revealed by the entanglement spectra in all cases.

\begin{figure*}[tbp]
\centering
\includegraphics[angle=0,scale=0.38]{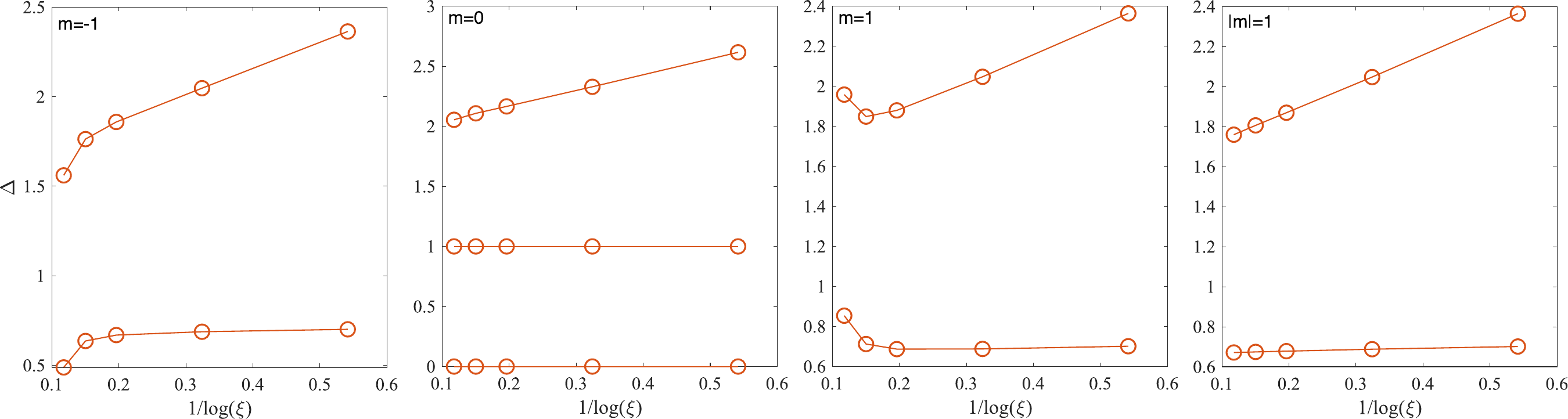}
  \caption{Entanglement spectra in charge $m$ sector from the $U(1)$-symmetric MPS for the XXZ spin chain Eq.~\ref{eq:xxz} at $g=0.5$. The last panel corresponds to the average of the $m=1$ and $m=-1$ results. The data has been normalized with the first energy gap in the charge $m=0$ sector.}
\label{fig_R_from_Es}
\end{figure*}

\subsection{Entanglement spectrum in XXZ spin chain}

\parR{Free compact boson on a cylinder.} The spin-half XXZ quantum spin chain 
\begin{equation}\label{eq:xxz}
    H = \sum_i \left( S^x_i S^x_{i+1} + S^y_i S^y_{i+1} + g\,S^z_i S^z_{i+1}\right) 
\end{equation}
has gapless ground state at $\vert g \vert < 1$. At low energy, it can be described by a $c=1$ compact boson CFT~\cite{Blumenhagen:2009zz}
\begin{equation}
 S = \frac{1}{4\pi}\int \partial \phi \, \bar{\partial}\phi   
\end{equation}
where $\phi(z,\bar{z}) = X + \bar{X}$ is compactfied on a circle $\phi \sim \phi + 2\pi R$. One can also introduce its dual field $\theta=X-\bar{X}$, also a compactfied scalar field $\theta \sim \theta + 4\pi/R$. The chiral component $X(z)$ can be expanded in terms of current operators $j_n$~\cite{Blumenhagen:2009zz}
\begin{equation}
    X(z) = x_0 - j_0\, i\,\mathrm{log} (z) + \sum_{n\neq0} i\, \frac{1}{n}\,j_n z^{-n},
\end{equation}
where the chiral zero mode satisfies $\left[x_0,\,j_0\right] = i$. When defined on a torus, the zero mode is quantized as $j_0 = \frac{m}{R} + \frac{eR}{2}$ and $\bar{j}_0 = \frac{m}{R} - \frac{eR}{2}$. The ground state can be labeled by these two charges $\vert m,e \rangle$. The modular invariant partition function is a sum over all these charge sectors
\begin{equation}
    Z = \sum_{e,m} Z_{0}\,q^{\frac{\left(m/R + eR/2\right)^2}{2}} \bar{q}^{\frac{\left(m/R - eR/2\right)^2}{2}}
\end{equation}
where $Z_0$ is the partition function of the non-compact boson (without the contribution of zero modes) and $q$ is the modular parameter. In a quantum theory, the energy spectrum has the form of
\begin{equation}
    E(e,m,n) \sim \frac{m^2}{R^2} + \frac{e^2\,R^2}{4} + n 
\end{equation}
where $n$ denotes non-negative integers.

When defined on an open cylinder (the space $\sigma$ direction is open), there are two different kinds of boundary conditions, the Neumann free boundary $\partial_\sigma \phi\vert_{b} = 0$ (or equivalently $\theta \vert_b=\theta_0 = x_0-\bar{x}_0$) and the Dirichlet fixed boundary $\phi\vert_b=\phi_0=x_0 + \bar{x}_0$. So the two boundary conditions are nothing but pinning down either the scalar field $\theta$ or $\phi$ at the boundary. Combining boundary conditions on both ends one can find 4 different cases. The two current algebras are reduced to a single one due to the existence of boundaries. The current operator satisfies $j_n=\bar{j}_n$ or $j_n = -\bar{j}_n$. Note that in the mixed boundary case $n \in \mathcal{Z}+\frac{1}{2}$ so that in the mode expansion of $j(z)$ there is no zero mode contribution. Solving the boundary conditions for the scalar fields $\phi$ and $\theta$, the zero modes $j_0$ read
\begin{equation}
    \begin{aligned}
        &j_0^{D,D} = \Delta \phi_0/2\pi + e\,R/2 \\
        &j_0^{N,N} = \Delta \theta_0/2\pi + m/R \\
        &j_0^{D,N} = j_0^{N,D} = 0,
    \end{aligned}
\end{equation}
where $\Delta \phi_0$ ($\Delta\theta_0$) is the difference of scalar field $\phi$ ( $\theta$ ) on the two boundaries, the zero mode equals zero in the mixed boundary BCFT since $n$ takes half-integers. One can also write down the energy spectrum for different boundary conditions
\begin{equation}
\begin{aligned}
    &E^{D,D} \sim \frac{\left( \Delta \phi_0/2\pi + e\,R/2 \right)^2}{2} + n \\
    &E^{N,N} \sim \frac{\left( \Delta \theta_0/2\pi + m/R \right)^2}{2} + n \\
    &E^{D,N} = E^{N,D} \sim n.
\end{aligned}
\end{equation}
One can find the energy spectrum does not depend on the radius $R$ in the mixed boundary conditions since there are no zero modes.

\parR{Boundary conditions in XXZ spin chain.} Through the technique of abelian bosonization one can identify the correspondence between lattice spin and field operators. In the following, we are going to show how to use lattice operators to impose boundary conditions. Here in our convention, we have the mapping between free fermion and compact boson $\psi^\dagger \sim \frac{1}{\sqrt{2\pi}} e^{iX(z)}$~\cite{giamarchi2003quantum}. Identifying the up or down state in the spin theory as an empty or occupied state in the hard-core boson, one can use the standard Jordan-Wigner transformation to build up the bosonization dictionary for the quantum spin chain
\begin{equation}
    \begin{aligned}
        S^z(x) &\sim - (-1)^x\, 2 \,\mathrm{cos}({\theta}R) - R \,\partial_x \theta \\
        S^-(x) &\sim (-1)^x \, e^{i\phi\,/2R} + e^{i\phi/2R} \mathrm{cos}({\theta}R) 
    \end{aligned}
\end{equation}
where ${\theta}$ is the dual field of $\phi$. Considering only the most relevant (the first term in the above dictionary) component, one can find that a boundary field along the $x-y$ plane or the $z$ direction to pin down the compact boson field $\phi$ or its dual field ${\theta}$ respectively. One can also shift a $1/2$ charge ($\phi$ or ${\theta}$) by reversing one of the boundary fields, which corresponds to introducing a flux term. One can also consider a dimerized perturbation $D(x) := (-1)^x \left[ S^x(x) S^x(x+a) + S^y(x) S^y(x+a) \right]$ to the spin chain. From abelian bosonization one finds 
\begin{equation}
     D(x) \sim A \ \mathrm{sin} ({\theta}R),
\end{equation}
where $A$ is a constant denoting the strength. It breaks the translation invariance ($T: \theta \rightarrow \theta + \pi/R$) explicitly, while the $x-y$ plane rotation symmetry ($R_a:\phi \rightarrow \phi + a$ where a being an arbitrary rotated phase) is preserved. Note that this operator is similar to the effect of the leading term in the $S^z(x)$, which can also be used to pin down the ${\theta}$ field at the boundary. This is not surprising: Compared with $S^z(x)$, there is a lattice constant shift in the composite operator which introduces an extra complex phase $e^{i k_f a}$ factor. Therefore $D(x)$ and $S^z(x)$ (the leading term) correspond to the imaginary and real fluctuation components respectively. Minimizing this term pins down $\theta$ and gives rise to two degenerate dimerization states ($A>0$ or $A<0$).

To verify these bosonization arguments, we consider a finite spin chain and put boundary fields at two boundaries of the model Eq.~\ref{eq:xxz}. In Fig.~\ref{fig_xxz-e} we show the result in the $\theta$ charge sector by putting boundary fields in the $z$ direction. The odd $N$ spin chain case corresponds to shifting a $1/2$ charge. In both cases, the low-energy spectra fit well with the BCFT predictions. Similarly, by putting a $x$ direction field at the boundary, we realize the BCFT in the $\phi$ charge sector, as shown in Fig.~\ref{fig_xxz-m}. By pinning the two boundaries in different directions, we can realize a mixed boundary BCFT, as shown in Fig.~\ref{fig_xxz-mix}, in which the energy spectra do not depend on the radius of the compact boson at all. These results confirm our discussion from abelian bosonization. 

\parR{Entanglement spectrum in XXZ spin chain.} In the study of the entanglement spectrum in the main text, the BCFTs realized in the MPS follows similarly. With $U(1)$ symmetry or not, the MPS states break either the translation or the spin rotation symmetry in the $XY$ plane. These correspond to pin down the scalar field $\theta$ or $\phi$ at the effective physical boundary respectively. Assuming the entanglement boundary to be free for the $\phi$ field (or equivalently fixed for $\theta$), we find the $U(1)$-symmetric MPS case correspond to a BCFT with a zero mode in the $\theta$ charge sector $j_0^{N,N}$. It is interesting that there are two different bonds in the MPS, in which the odd ones have integer charges and the even bonds have half-integer charges. The latter case can either be understood as fractionalization of the $U(1)$ symmetry or the field $\theta$ at the entanglement boundary has been shifted half of its periodicity. The generic-MPS case realize an mixed boundary BCFT in which the zero mode $j_0^{N,D}$ disappears. The spectrum does not depend on the radius $R$ or $g$ at all. 

As an application of our study of the entanglement spectrum, one can use the dependence of entanglement spectrum on the radius to calculate $R$, hence the Luttinger parameter. For example, from the charge $m=1$ first level spectrum $\Delta$, one can extract $R$ according to the relation $\Delta=4/R^2$. One subtlety arises here in our simulation with only $U(1)$ symmetry without considering the charge conjugation $\mathbb{C}$. In Fig.~\ref{fig_R_from_Es}, we show the scaling of the entanglement spectrum with respect to $1/\mathrm{log}(\xi)$ at $g=0.5$. At large bond dimension we find the particle-hole symmetry can be spontaneously broken which makes the entanglement spectrum deviates from the $1/\mathrm{log}(\xi)$ scaling. Nevertheless, the average of the positive and negative charge restores the correct leading correction predicted by BCFT. The radius $R$ is fitted to be $R=1.736$ which is quite close to the theoretical value $\sqrt{3} \approx 1.732$. Finite scaling of entanglement spectrum thus can be used to calculate the Luttinger parameter in other models with $U(1)$ symmetry described by $c=1$ compact boson CFT at the low energy. 

\parR{The nearest neighbor RVB wave function on the square lattice.}
\begin{figure}[tbp]
\centering
\includegraphics[angle=0,scale=0.42]{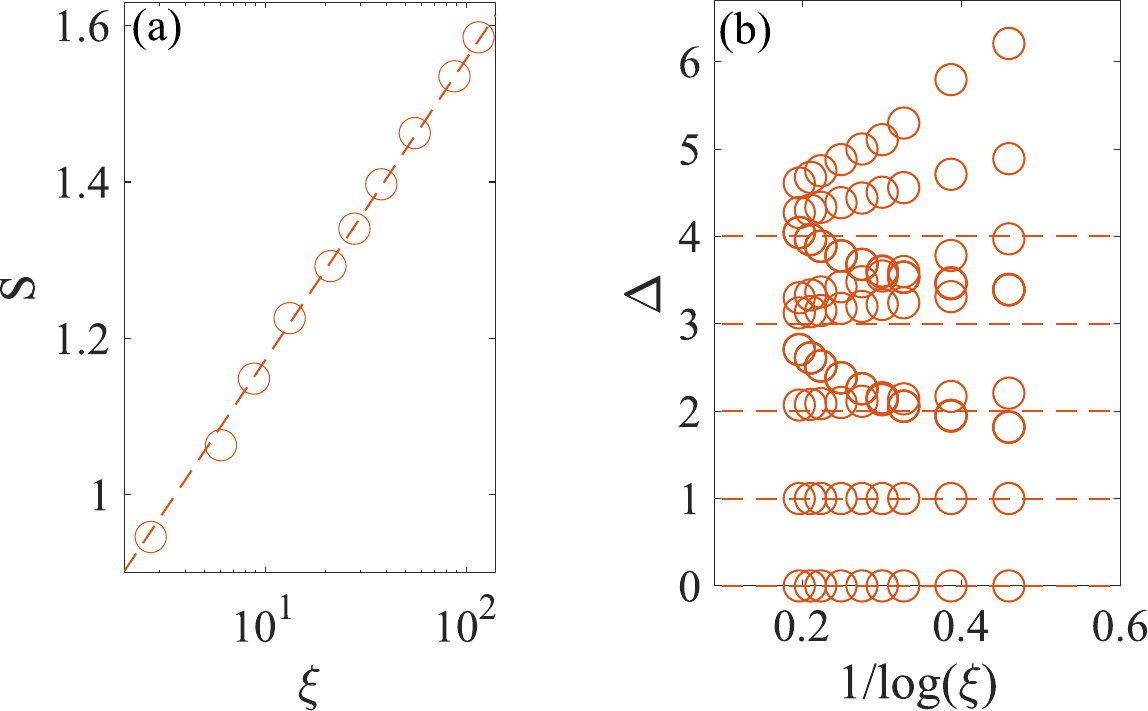}
  \caption{Entanglement properties of the boundary MPS for the nearest neighbor RVB wave function. (a) Finite correlation length scaling of the entanglement entropy. The central charge is found to be $c=1.003$. (b) Low-lying entanglement spectrum.}
\label{fig_rvb}
\end{figure}
It is interesting that our analysis also applies to a large class of two-dimensional wavefunctions represented by infinite projected entangled pair states (PEPS). These systems are of the Rokhsar-Kivelson type, or more generally the so-called conformal quantum critical points, whose low energy modes have a dispersion $\omega \sim k^2$. These theories have a dynamical exponent $z=2$ and possess conformal symmetry in space. Their equal-time properties can be understood using two-dimensional CFT~\cite{ARDONNE2004493}. In such theories, the ground state overlap is equivalent to a two-dimensional critical statistical model. The PEPS transfer matrix serves the role of a critical Hamiltonian in the quantum spin chain. This allows us to use MPS and entanglement scaling to study the entanglement spectrum contained in the boundary MPS as the leading eigenvector of the transfer matrix. Note that the entanglement discussed here is defined in the boundary MPS rather than in the PEPS.

Here we take the nearest neighbor RVB state~\cite{anderson1987resonating} on the square lattice as an example. The RVB state is a class of many-body wave functions composed of a superposition of singlets on bonds. It is believed these states are useful in the description of quantum spin liquids, i.e. the topologically ordered states in frustrated quantum spin models. The nearest neighbor RVB on the square lattice is one of the simplest examples of realizing a quantum spin liquid. It can be represented by a $D=3$ PEPS~\cite{frank_2006}. The local tensor in the PEPS is defined by combining the projector
\begin{equation}
    \begin{aligned}
        P = \,&\vert 0 \rangle \left\{ ( 0222\vert + (2022\vert + (2202\vert + (2220\vert\right\} + \\
        & \vert 1 \rangle \left\{ ( 1222\vert + (2122\vert + (2212\vert + (2221\vert\right\}
    \end{aligned} 
\end{equation}
and the singlet state ($\vert 01) - \vert 10)$) on the bonds~\cite{frank_2006}. This state is gapless and the ground state overlap is equivalent to a two-dimensional partition function that can be described by a $c=1$ free compact boson CFT~\cite{dreyer2020robustness}. In Fig.~\ref{fig_rvb}, we show the entanglement properties in the boundary MPS. The entanglement entropy satisfies a perfect scaling form $S \sim \frac{c}{6} \log (\xi) + \mathrm{constant}$ \cite{Pollmann2009}, with the central charge is fitted to be $c=1.003$, which is quite close to the exact one. Without symmetry constraint, our theory predicts that the entanglement spectrum should follow a mixed boundary BCFT as in the XXZ quantum spin chain case. The numerical data again confirm our analysis as shown in Fig.~\ref{fig_rvb} (b). The low-lying spectrum approaches to integers and does not depend on the radius of the free compact boson.

\end{document}